\documentclass[10pt]{article}

\usepackage{amsmath}
\usepackage{amssymb}

\usepackage{soul}

\usepackage{graphicx}

\usepackage{cite}

\usepackage{color} 

\usepackage{subfigure}
\usepackage{bm}            

\usepackage[labelfont=bf,labelsep=period,justification=raggedright]{caption}

\makeatletter
\renewcommand{\@biblabel}[1]{\quad#1.}
\makeatother

\date{}

\begin{document}

\begin{flushleft}
{\Large
\textbf{Interplay between telecommunications and face-to-face interactions - a study using mobile phone data}
}
\\
Francesco Calabrese$^{1,2,\ast}$, 
Zbigniew Smoreda$^{3}$, 
Vincent D. Blondel$^{4,5}$,
Carlo Ratti$^{2}$
\\
\bf{1} IBM Research, Dublin, Ireland
\\
\bf{2} Senseable City Laboratory, Massachusetts Institute of technology, Cambridge, Massachusetts, United States of America 
\\
\bf{3} Sociology and Economics of Networks and Services Department, Orange Labs, Issy-les-Moulineaux, France
\\
\bf{4} LIDS, Massachusetts Institute of technology, Cambridge, Massachusetts, United States of America 
\\
\bf{5} Department of Mathematical Engineering, Universit\'e catholique de Louvain, Louvain-la-Neuve, Belgium
\\
$\ast$ E-mail: Corresponding fcalabre@ie.ibm.com
\end{flushleft}

\section*{Abstract}
In this study we analyze one year of anonymized telecommunications data for over one million customers from a large European cellphone operator, and we investigate the relationship between people's calls and their physical location. We discover that more than 90\% of users who have called each other have also shared the same space (cell tower), even if they live far apart. Moreover, we find that close to 70\% of users who call each other frequently (at least once per month on average) have shared the same space at the same time - an instance that we call co-location. Co-locations appear indicative of coordination calls, which occur just before face-to-face meetings. Their number is highly predictable based on the amount of calls between two users and the distance between their home locations - suggesting a new way to quantify the interplay between telecommunications and face-to-face interactions.

\section*{Introduction}

The interplay between telecommunications, travel and face-to-face meetings is an unresolved puzzle. In some cases it has been suggested that telecommunications may be a substitute for physical interaction \cite{Albertson} - an idea that gained traction during the nineties and the rapid expansion of the Internet \cite{cairncross,mitchell}. In other cases conflicting hypotheses have been made, including those of a complementary \cite{Mokhtarian,mok2010}, neutral \cite{choo} or reinforcing \cite{sasaki2010} effect.
Recently, social networks have been identified as possible predictors of travel behavior, as well as the possible decision to telecommute \cite{Salomon199817,paez2007}. Social interaction has thus been integrated in activity-travel models \cite{arentze2008}, in addition to the existing categories of travel such as commuting, leisure and business. Furthermore, researchers such as Urry and others \cite{urry,larsen_book,sheller} have argued that flows and meetings of people produce small worlds, which require connections and meeting places - a phenomenon which is also known as the new mobilities paradigm. 

This study aims to provide a new perspective into the relationship between telecommunicating people and their physical locations through an assesment of anonymized Call Detail Records (CDRs). 
CDRs show great promise for academic research: they have recently been used to explore human communications \cite{onnela2007sat, eaglepnas}, the geography of social networks \cite{lambiotte2008gdm,krings2009}, urban dynamics \cite{CalabreseReades2007}, and human mobility patterns \cite{Gonzalez:2008uq,Wang:2009fk,ChaomingSong02192010,pervasive2010}.
In this paper we use them for the first time to study the relationship between the telecommunications patterns of any two people and their physical locations.

\section*{Results}

\label{sec:coloc}

We use a large anonymized dataset of over one million mobile phone users, which was gathered in Portugal over a twelve month period between 2006 and 2007. To safeguard personal privacy, individual phone numbers were anonymized by the operator before leaving storage facilities, and they were identified with a security ID (hash code). Each entry in the dataset has a CDR, which consists of the following information: timestamp, callerÕs ID, calleeÕs ID, call duration, callerÕs cell tower ID, and callee's cell tower ID. This metadata on each call allows us to study both the mobile social interaction as well as the physical location of the users within the dataset. Notably, the dataset does not contain information regarding text messages (SMS) or data usage (internet). More details about the dataset can be found in Text S1.

In this study, we start with the initial dataset and look at all communications between pairs of users, together with their locations at call time. As we are interested in comparing people locations, we discard users for which we do not have enough samples. We use two subsets: D1, which contains all reciprocal communications between the top 100,000 callers; and D2, which contains 10,000 pairs from D1, sampled at different home distances to ensure the same home distances distribution found in D1 (see Text S2). In the sequel, we use D2 in cases where computational complexity limits the use of a larger set.

We discover that at least 93\% of users in D1 who reciprocally call each other, have at least once shared the same cell tower area in one year.
The percentage decreases slightly as the distances between their homes decreases, but the value is still above 90\% for users living 100 km apart (see Figure 1).
It appears that almost all remote communications are associated with being physically sharing space. 
It may also be noted that we are underestimating the percentage as our data is only based on locations at call time, so users might have also shared space without this being recorded in our data.
Results are consistent with what was recently found analyzing spatio-temporal coincidences in a geo-tagged pictures database to infer social ties \cite{crandall2010}.

If we also consider the temporal component, we can look at how often and where users are sharing the same space at the same time.
We restrict our attention to the case when two users call each other using the same cell tower.
This scenario is based on the hypothesis that they are calling each other to coordinate to meet in a nearby area, also called ``coordination knot'' \cite{smoreda_coordination_call2008}.
Of course, two people living or working close by could also call each other very often without physically meeting. So, we excluded users living or working in the same cell tower area, estimated as described in Text S2. 
\textit{We define a co-location event between two users (who live and work in distinct locations) as a call between the users while they are connected to the same cellphone tower. Each co-location is characterized by a specific time and place.}
Based on this definition, we characterize the spatio-temporal features of co-location events, to see whether they represent a reasonable subset of actual face-to-face meetings between users.

Starting with the larger subset (D1), we analyze the relationship between calling activity and  user's locations.
Among the pairs of communicating users, 400,000 cases have two users calling each other while in the same cell tower area, 350,000 of which have distinct home and work locations.
Interestingly, 38.33\% of the communicating users co-locate at least once during the period examined. When stronger relationships are considered (users who call on average at least once per month) the percentage increases to 69.41\%.

Call duration appears to increase with the homes distance between users (see black line in Figure 2). Calls that occur between co-located people (red line) have a shorter average call duration, suggesting that people who co-locate call each other briefly to coordinate the exact meeting place and time.

We also find that the number of calls between two users increases just before and after their co-location (Figure 3).
The probability is rather constant in the interval, with two peaks around 0 and 1 (consecutive co-location events).
The presence of these peaks suggests that the considered events (co-locations) represent a reasonable proxy for face-to-face meetings.
In particular, a peak of calls just before the co-location event, suggests that the two people are talking on the phone to arrange a meeting, in line with what is hypothesized in \cite{lambiotte2008gdm,smoreda_coordination_call2008}. The peak right after the co-location event might be explained by a follow up call after the meeting.

We analyze the features of co-location places and compared it with geographical and communication differences between users. We define $d_1(l)$ and $d_2(l)$ as the distances traveled by two users at every co-location event $l=1,\dots,m$, and compute three measures of comparison:
\begin{enumerate}
\item
The median ratio between the shortest and longest distance at co-location time:
\[
r_d=median_{l} \frac{min \{d_1(l), d_2(l)\}}{max \{d_1(l), d_2(l)\}}.
\]
\item
The fraction of times user $1$ travels less than its peer:
\[
r_{t1} = 1/m \sum_{l=1}^m g(d_2(l)-d_1(l)),
\]
where:
\[
g(x)=\left\{
\begin{array}{cc}
1  &  if \; x>0  \\
 0 &  if \; x=0
\end{array}
  \right. .
\]
\item
The fraction of times one of the users travels less than the peer:
\[
r_t = min \{ r_{t1}, r_{t2}  \}.
\]
\end{enumerate}

The first measure $r_d$ allows a comparison to be made between the lengths of the two users' trips. On the D2 subset, we find on average $r_d=0.3$, i.e. one user travels about 3 times less than the other one.
Due to the asymmetric behavior in the length of trips, we question whether the shorter trips are always taken by the same user, or if the two users share the short trips. The third measure $r_l$ allows an evaluation of the asymmetry at the pair level, showing an average of $0.06$. This suggests that in 94\% of the selected pairs, there is one user who constantly travels less than its peers.
The second measure $r_{l1}$ is a directed measure and is computed to see whether geographical and communication differences allow the user that travels less to be predicted. Text S3 reports how these measures vary with homes distance, population density, normalized tie strength and call direction. In particular we find that as users' homes distance increases, co-locations occur in a place that is closer to one of the users. Moreover, the more the normalized tie strength differs between users, the more the co-locations occur in places close to one of them.

Our definition of distance $d$ is based on the Euclidean distance between home and co-location places. Two limitations arises from this choice:
1) the Euclidean distance does not take into account the real path taken by a person; 2) the person might not travel directly from home but the origin of the trip to the co-location place could be different. However, as we are interested in the relative distances traveled by the two peers, we can assume that both limitations affect the two measures in a similar manner, thus limiting the potential bias.

We evaluate the relationship between the home locations' distance and the number of co-locations between users. Figure 4(a) shows the average number of co-locations, which decreases with distance.
The result is consistent with what was found in \cite{larsen,larsen_book, Wheeler, Carrasco} using data from surveys.
If we compare this decrease with the one of phone calls, and total call times (see Figure 4(a)) we find different decays with distance. Total call time is the least affected by distance (slope -0.04), followed by the number of calls (slope -0.07). In contrast with this, the number of co-locations is strongly affected by distance (slope -0.14). 
Even if we consider a broader definition of co-location, in which two users are considered co-located in the same cell tower if they happen to make a phone call (not necessarily to each other) from the same cell tower area within one hour, we still find a similar decreasing trend, as shown in Figure 4(b) computed for the D2 subset.
The results are consistent with those from the analysis of fixed phone data combined with interviews showing the effect of distance on call duration and frequency of meetings \cite{Smoreda2006,Smoreda}.

The number of calls has a strong influence on the number of co-locations, suggesting that the more people call each other, the more they co-locate (see Figure 5).
As there appears to be a clear relationship between call patterns, distance and co-locations, we tried to built a predictor of the number of co-locations, starting from a measure of interaction (number of calls) and the geographical distance between users' home, obtaining $r^2=0.61$ with the model (Figure 6): 
\[
\#colocations=0.92 \frac{\#calls^{0.60}}{distance^{0.08}}. 
\]
This result suggests that geography and telecommunication interactions account for 61\% of variations in the number of co-locations (see also Text S4). 
This is consistent during the one year time frame under analysis, as reported in Text S5.
The exponent $0.60$ for the $\#calls$ reveals the correlation between an increase in the number of calls and an increase in the number of co-locations.
This result suggests that telecommunications might play a complementary role in facilitating face-to-face interactions, supporting the observations found in other studies \cite{Mokhtarian,mok2010}.

\medskip

\section*{Discussion}
In this study we analyze one year of telecommunications data from a large European cellphone operator to investigate the relationship between people's calls and their physical location.

We discover that more than 90\% of users who called each other have also shared the same space (cell tower), even if they live far apart. Moreover, we find that 69\% of users who call each other frequently (at least once per month on average) have shared the same space at the same time - an instance that we call co-location. 
Co-locations appear highly indicative of coordination calls occurring just before face-to-face meetings. We are able to predict 61\% of variations in the number of co-locations from the number of calls, and users' homes distance. In particular, as the distance between homes increases, the expected number of co-locations decreases.

We also characterize the co-location places in terms of distance from the home locations.
As the users' homes distance increases, co-locations occur in a place that is closer to one of the users.
In more than 90\% of the cases, co-locations take place in an area that is closer to the same user of the pair (there is low reciprocity in the travel distance covered). Telecommunication strength helps predict which person of the pair travels less.

We believe that the above results suggest new ways to use CDRs to investigate the old conundrum of the interplay between telecommunications, travel and face-to-face meetings - with applications in the social sciences, urban planning and transportation studies.

\section*{Acknowledgments}

The authors thank Dima Ayyash, Dominik Dahlem, Santi Phithakkitnukoon and Prudence Robinson for their feedback, and Orange Labs, IBM Research, the National Science Foundation, the AT\&T Foundation, the MIT SMART program, GE, Audi Volkswagen, SNCF, ENEL and the members of the MIT Senseable City Lab Consortium for supporting the research.

\bibliography{bibfile}

\begin{thebibliography}{10}
\providecommand{\url}[1]{\texttt{#1}}
\providecommand{\urlprefix}{URL }
\expandafter\ifx\csname urlstyle\endcsname\relax
  \providecommand{\doi}[1]{doi:\discretionary{}{}{}#1}\else
  \providecommand{\doi}{doi:\discretionary{}{}{}\begingroup
  \urlstyle{rm}\Url}\fi
\providecommand{\bibAnnoteFile}[1]{%
  \IfFileExists{#1}{\begin{quotation}\noindent\textsc{Key:} #1\\
  \textsc{Annotation:}\ \input{#1}\end{quotation}}{}}
\providecommand{\bibAnnote}[2]{%
  \begin{quotation}\noindent\textsc{Key:} #1\\
  \textsc{Annotation:}\ #2\end{quotation}}
\providecommand{\eprint}[2][]{\url{#2}}

\bibitem{Albertson}
Albertson LA (1977) Telecommunications as a travel substitute: Some
  psychological, organizational, and social aspects.
\newblock Journal of Communication 27: 32-43.
\bibAnnoteFile{Albertson}

\bibitem{cairncross}
Cairncross F (1997) The death of distance.
\newblock Harvard Business School Press.
\bibAnnoteFile{cairncross}

\bibitem{mitchell}
Mitchell WJ (1996) City of bits: space, place, and the infobahn.
\newblock MIT Press.
\bibAnnoteFile{mitchell}

\bibitem{Mokhtarian}
Mokhtarian PL (2003) Telecommunications and travel: The case for
  complementarity.
\newblock Journal of Industrial Ecology 6: 43-57.
\bibAnnoteFile{Mokhtarian}

\bibitem{mok2010}
Mok D, Wellman B, Carrasco J (2010) Does distance matter in the age of the
  internet?
\newblock Urban Studies 47: 2747-2783.
\bibAnnoteFile{mok2010}

\bibitem{choo}
Choo S, Lee T, Mokhtarian PL (2010) Do transportation and communications tend
  to be substitutes, complements, or neither? u.s. consumer expenditures
  perspective, 1984-2002.
\newblock Transportation Research Record : 121-132.
\bibAnnoteFile{choo}

\bibitem{sasaki2010}
Sasaki K, Nishii K (2010) Measurement of intention to travel: Considering the
  effect of telecommunications on trips.
\newblock Transportation Research Part C 18: 36-44.
\bibAnnoteFile{sasaki2010}

\bibitem{Salomon199817}
Salomon I (1998) Technological change and social forecasting: the case of
  telecommuting as a travel substitute.
\newblock Transportation Research Part C: Emerging Technologies 6: 17-45.
\bibAnnoteFile{Salomon199817}

\bibitem{paez2007}
Paez A, Scott D (2007) Social influence on travel behavior: a simulation
  example of the decision to telecommute.
\newblock Environment and Planning A 39: 647-665.
\bibAnnoteFile{paez2007}

\bibitem{arentze2008}
Arentze T, Timmermans H (2008) Social networks, social interactions, and
  activity-travel behavior: a framework for microsimulation.
\newblock Environment and Planning B 35: 1012-1027.
\bibAnnoteFile{arentze2008}

\bibitem{urry}
Urry J (1999) Sociology beyond societies: mobilities for the 21st century.
\newblock Routledge, London.
\bibAnnoteFile{urry}

\bibitem{larsen_book}
Larsen J, Urry J, Axhausen K (2006) Mobilities, networks, geographies.
\newblock Ashgate, Aldershot.
\bibAnnoteFile{larsen_book}

\bibitem{sheller}
Sheller M, Urry J (2006) The new mobilities paradigm.
\newblock Environment and Planning A 38: 207-226.
\bibAnnoteFile{sheller}

\bibitem{onnela2007sat}
Onnela J, Saramaki J, Hyvonen J, Szabo G, Lazer D, et~al. (2007) Structure and
  tie strengths in mobile communication networks.
\newblock Proceedings of the National Academy of Sciences 104: 7332.
\bibAnnoteFile{onnela2007sat}

\bibitem{eaglepnas}
Eagle N, Pentland AS, Lazer D (2009) Inferring friendship network structure by
  using mobile phone data.
\newblock Proceedings of the National Academy of Sciences 106: 15274--15278.
\bibAnnoteFile{eaglepnas}

\bibitem{lambiotte2008gdm}
Lambiotte R, Blondel V, de~Kerchove C, Huens E, Prieur C, et~al. (2008)
  {Geographical dispersal of mobile communication networks}.
\newblock Physica A: Statistical Mechanics and its Applications 387:
  5317--5325.
\bibAnnoteFile{lambiotte2008gdm}

\bibitem{krings2009}
Krings G, Calabrese F, Ratti C, Blondel V (2009) A gravity model for inter-city
  telephone communication networks.
\newblock {Journal of Statistical Mechanics: Theory and Experiment} L07003.
\bibAnnoteFile{krings2009}

\bibitem{CalabreseReades2007}
Reades J, Calabrese F, Sevtsuk A, Ratti C (2007) Cellular census: Explorations
  in urban data collection.
\newblock IEEE Pervasive Computing 6: 30-38.
\bibAnnoteFile{CalabreseReades2007}

\bibitem{Gonzalez:2008uq}
Gonzalez M, Hidalgo C, Barabasi AL (2008) Understanding individual human
  mobility patterns.
\newblock Nature 453: 779--782.
\bibAnnoteFile{Gonzalez:2008uq}

\bibitem{Wang:2009fk}
Wang P, Gonzalez M, Hidalgo C, Barabasi AL (2009) Understanding the spreading
  patterns of mobile phone viruses.
\newblock Science 324: 1071--1076.
\bibAnnoteFile{Wang:2009fk}

\bibitem{ChaomingSong02192010}
Song C, Qu Z, Blumm N, Barabasi AL (2010) {Limits of Predictability in Human
  Mobility}.
\newblock Science 327: 1018-1021.
\bibAnnoteFile{ChaomingSong02192010}

\bibitem{pervasive2010}
Calabrese F, Pereira F, DiLorenzo G, Liu L (2010) The geography of taste:
  analyzing cell-phone mobility and social events.
\newblock In: International Conference on Pervasive Computing.
\bibAnnoteFile{pervasive2010}

\bibitem{crandall2010}
Crandall D, Backstrom L, Cosley D, Suri A, Huttenlocher D, et~al. (2010)
  Inferring social ties from geographic coincidences.
\newblock Proceedings of the National Academy of Sciences 107: 22436-22441.
\bibAnnoteFile{crandall2010}

\bibitem{smoreda_coordination_call2008}
Diminescu D, Licoppe C, Smoreda Z, Ziemlicki C (2008) The reconstruction of
  space and time. Mobile communication practices, New Crunswick and London:
  Transaction Publishers, chapter Tailing untethered mobile users: Studying
  urban mobilities and communication practices.
\bibAnnoteFile{smoreda_coordination_call2008}

\bibitem{larsen}
Larsen J, Axhausen K, Urry J (2006) Geographies of social networks: Meetings,
  travels and communications.
\newblock Mobilities 1: 261-283.
\bibAnnoteFile{larsen}

\bibitem{Wheeler}
Wheeler J, Stutz F (1971) Spatial dimensions of urban social travel.
\newblock Annuals of the Association of American Geographers 61: 371 - 386.
\bibAnnoteFile{Wheeler}

\bibitem{Carrasco}
Carrasco J, Miller E, Wellman B (2008) How far and with whom do people
  socialize?: Empirical evidence about distance between social network members.
\newblock Transportation Research Record: Journal of the Transportation
  Research Board 2076: 114-122.
\bibAnnoteFile{Carrasco}

\bibitem{Smoreda2006}
Licoppe C, Smoreda Z (2006) Computers, Phones, and the Internet : Domesticating
  Information Technology, Oxford University Press, chapter Rhythms and ties:
  towards a pragmatics of technologically-mediated sociability.
\bibAnnoteFile{Smoreda2006}

\bibitem{Smoreda}
Licoppe C, Smoreda Z (2000) Liens sociaux et r{\'e}gulations domestiques dans
  l'usage du t{\'e}l{\'e}phone. de l'analyse quantitative de la dur{\'e}e des
  conversations {\`a} l'examen des interactions.
\newblock R{\'e}seaux 18: 253--276.
\bibAnnoteFile{Smoreda}

\end{thebibliography}

\section*{Supporting Information Legends}

S1 Dataset\\
S2 Home and work location determination\\
S3 Co-location places, geography and communication strength\\
S4 Statistical analysis\\
S5 Relationship between co-locations and calls over time

\begin{figure}[h]
\centering
\includegraphics[width=0.6\columnwidth]{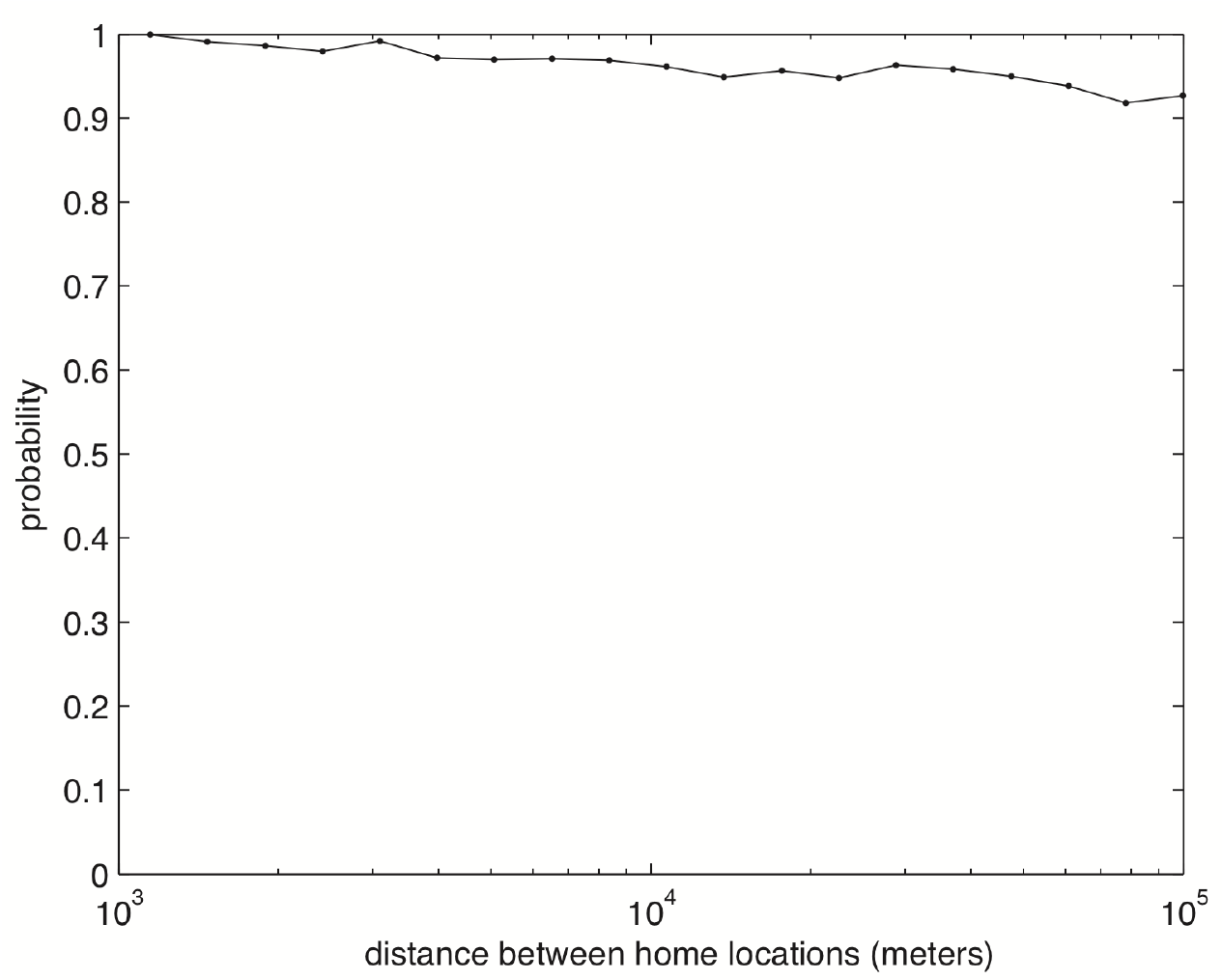}
\caption{Probability that two reciprocal calling users have shared the same cell tower area during a 1 year time (D1 subset).}
\end{figure}

\begin{figure}[h]
\centering
\includegraphics[width=0.6\columnwidth]{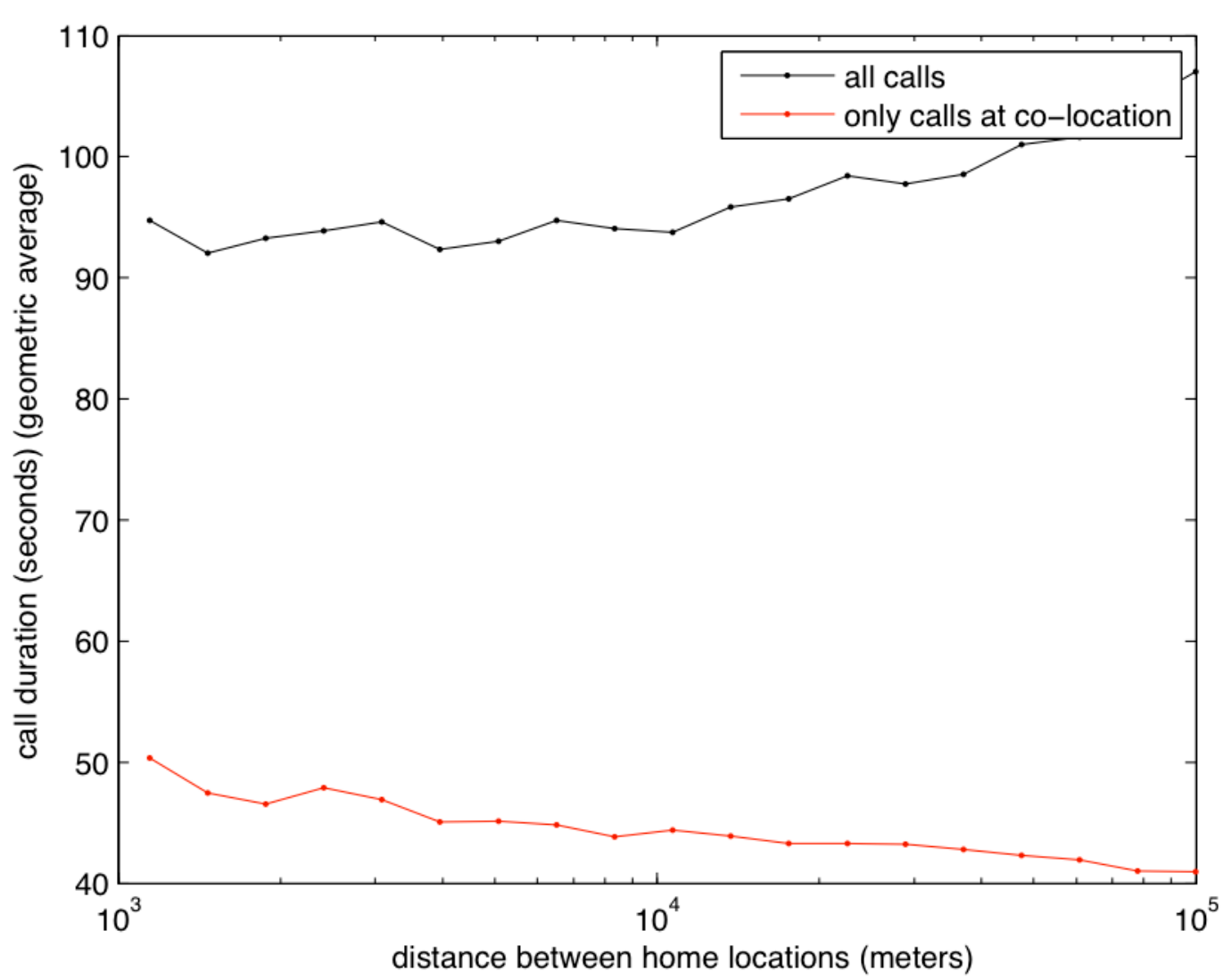}
\caption{Average length of a call as a function of the users' home distances (D1 subset).}
\end{figure}

\begin{figure}[h]
\centering
\includegraphics[width=0.6\columnwidth]{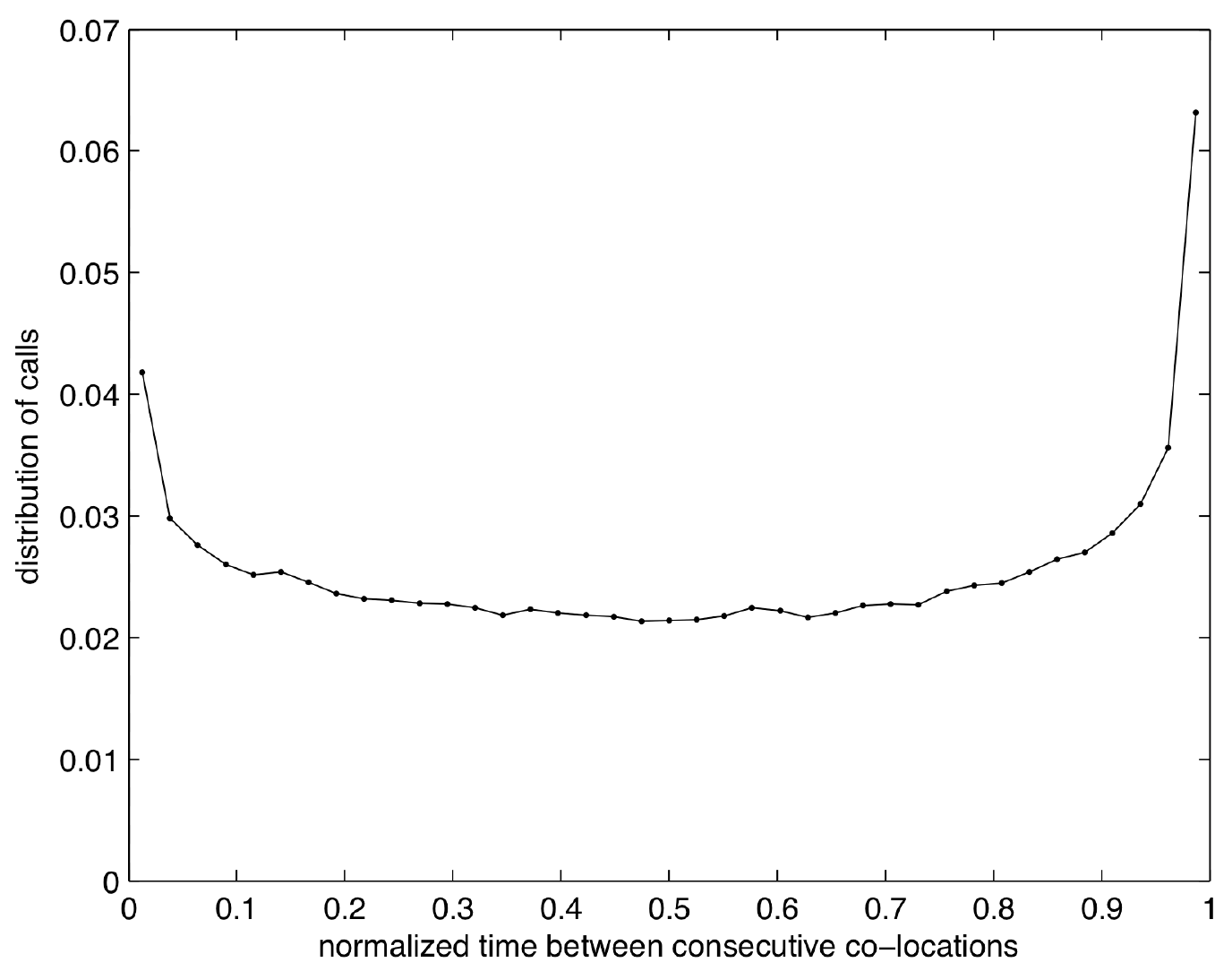}
\caption{Number of calls between consecutive co-locations. Call times have been normalized to the range of 0 to 1 (D2 subset).}
\end{figure}

\begin{figure}[h]
\centering
\includegraphics[width=\columnwidth]{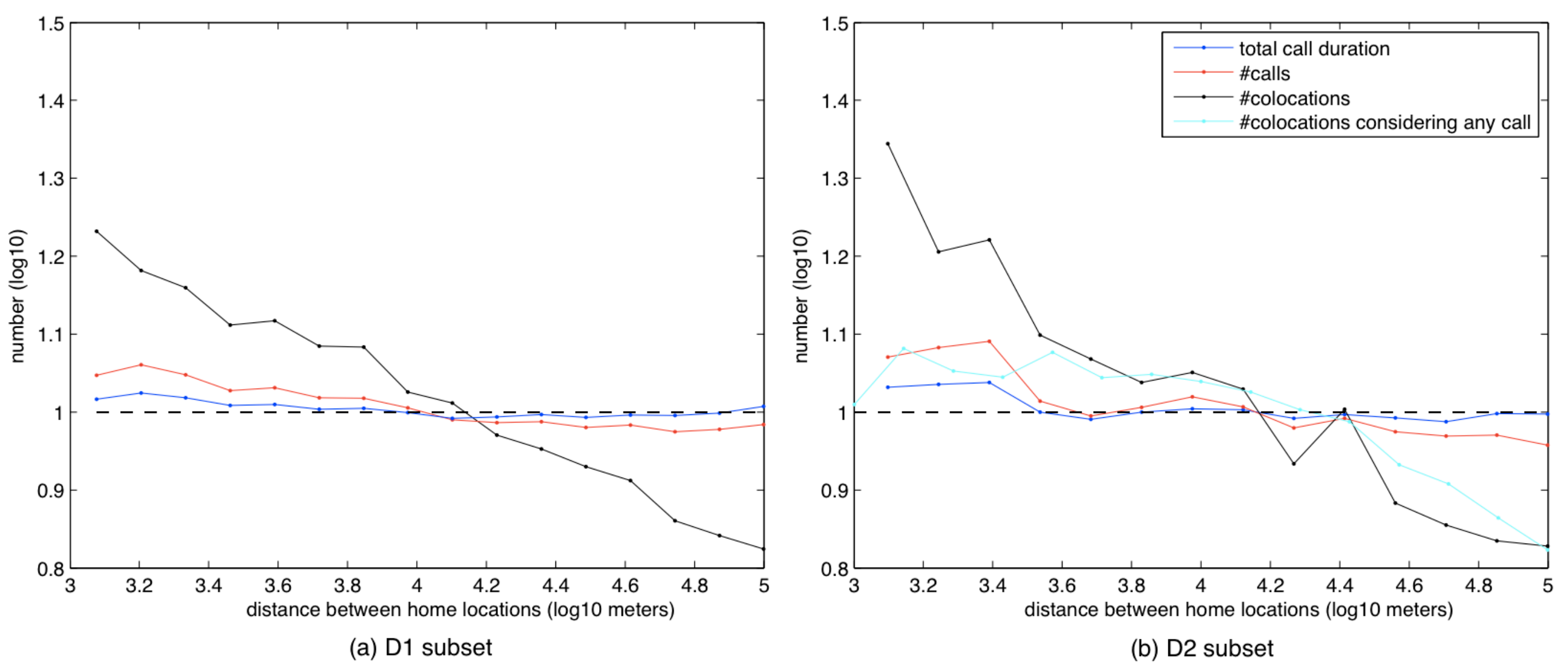}
\caption{Normalized number of co-locations, calls, total call duration as function of users' home distances.}
\end{figure}

\begin{figure}[h]
\centering
\includegraphics[width=0.6\columnwidth]{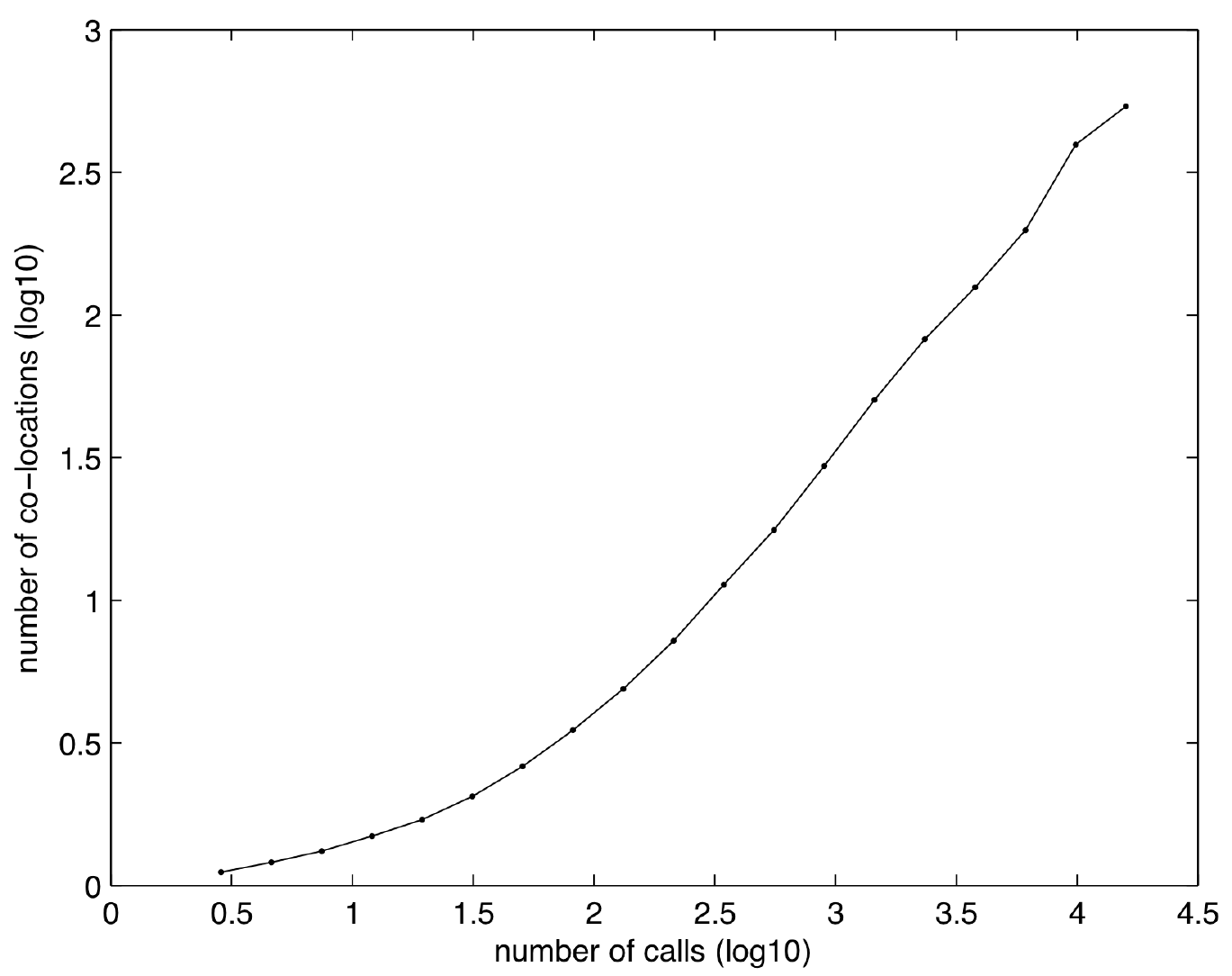}
\caption{Average number of co-locations as a function of the number of calls (D1 subset).}
\end{figure}

\begin{figure}[h]
\centering
\includegraphics[width=0.6\columnwidth]{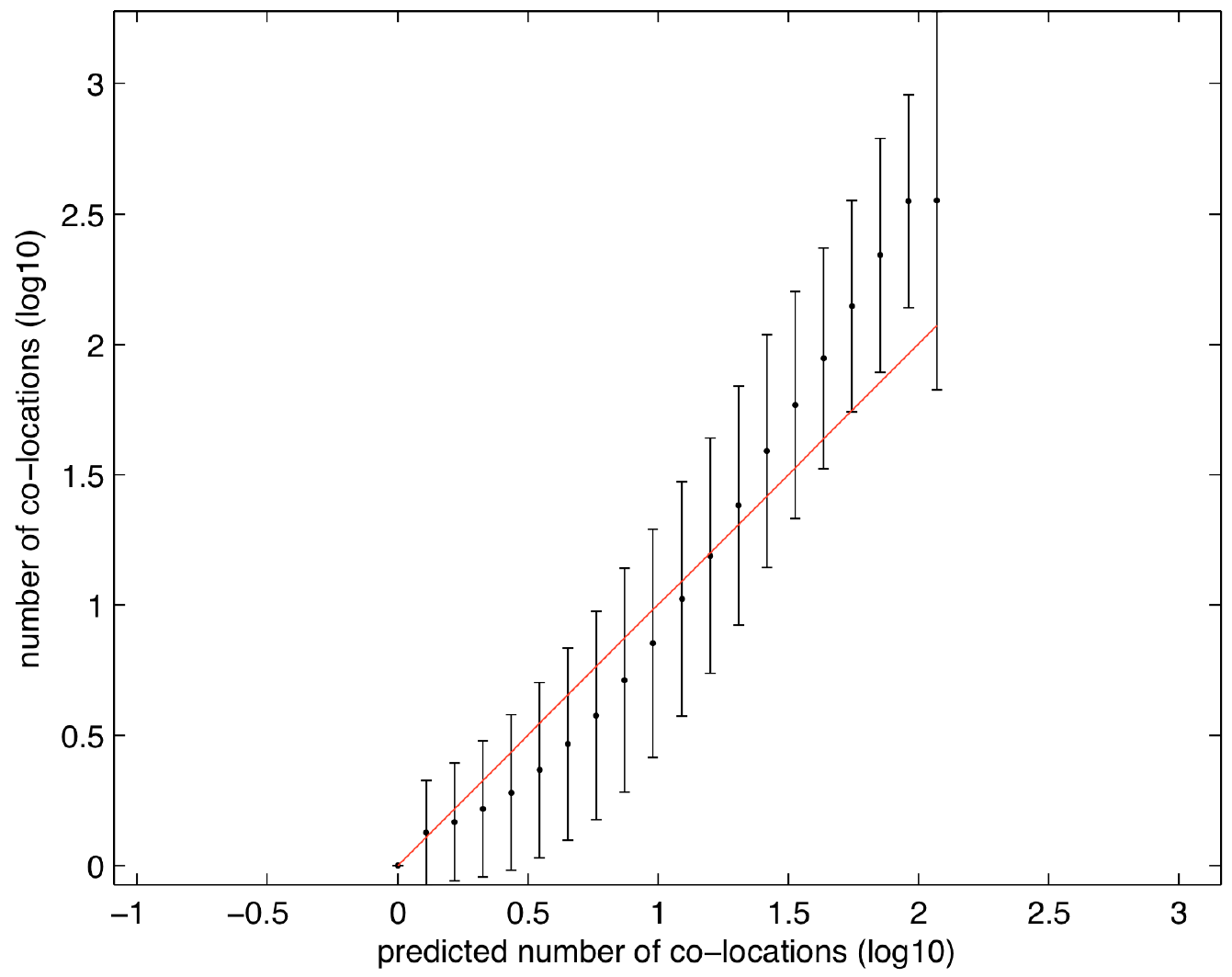}
\caption{A prediction of the number of co-locations. Error bars represent the standard deviations (D1 subset).}
\end{figure}

\end{document}